\documentclass[manuscript, longauth, twocolumn]{aa_longauth}
\usepackage[varg]{txfonts}

\usepackage{graphicx}
\usepackage{amssymb, amsmath} 
\usepackage{color}
\usepackage[usenames,dvipsnames,table]{xcolor}
\bibpunct{(}{)}{;}{a}{}{,} 
\usepackage{multirow}
\usepackage[left]{lineno}

\makeatletter
\renewcommand*{\@fnsymbol}[1]{\ifcase#1\or*\or$\dagger$\or$\ddagger$\or**\or$\dagger\dagger$\or$\ddagger\ddagger$\fi}
\makeatother

\newcommand{\modifPol}[1] {\textcolor{black}{#1}}
\newcommand{\modifAlicia}[1] {\textcolor{black}{#1}}
\newcommand{\modifPere}[1] {\textcolor{black}{#1}}

\def \deg{^\circ}

\def \hcm {\hbox {\ifmmode $ atom cm$^{-2}\else atom cm$^{-2}$\fi}}

\def\j11{\mbox{\object{IGR~J11014-6103}}}

\def \apjl {ApJL}
\def \apjs {ApJS}
\def \aap {A\&A}
 
\newcommand{\ssbin}{SS~433}

\begin{document}


\title{\modifAlicia{Constraints on particle acceleration in SS433/W50 \break from MAGIC and H.E.S.S. observations}}
\titlerunning{SS~433 VHE observations}
\authorrunning{MAGIC and H.E.S.S. Collaborations}

\author{{\textbf{MAGIC Collaboration}:
M.~L.~Ahnen\inst{1} \and
S.~Ansoldi\inst{2,}\inst{25} \and
L.~A.~Antonelli\inst{3} \and
C.~Arcaro\inst{4} \and
A.~Babi\'c\inst{5} \and
B.~Banerjee\inst{6} \and
P.~Bangale\inst{7} \and
U.~Barres de Almeida\inst{7,}\inst{26} \and
J.~A.~Barrio\inst{8} \and
J.~Becerra Gonz\'alez\inst{9,}\inst{10,}\inst{27,}\inst{28}
% ...
%
}}
\institute { ETH Zurich, CH-8093 Zurich, Switzerland
\and Universit\`a di Udine, and INFN Trieste, I-33100 Udine, Italy
\and INAF National Institute for Astrophysics, I-00136 Rome, Italy
\and Universit\`a di Padova and INFN, I-35131 Padova, Italy
\and Croatian MAGIC Consortium, Rudjer Boskovic Institute, University of Rijeka, University of Split - FESB, University of Zagreb - FER, University of Osijek,Croatia
\and Saha Institute of Nuclear Physics, 1/AF Bidhannagar, Salt Lake, Sector-1, Kolkata 700064, India
\and Max-Planck-Institut f\"ur Physik, D-80805 M\"unchen, Germany
\and Universidad Complutense, E-28040 Madrid, Spain
\and Inst. de Astrof\'isica de Canarias, E-38200 La Laguna, Tenerife, Spain
\and Universidad de La Laguna, Dpto. Astrof\'isica, E-38206 La Laguna, Tenerife, Spain
\and University of \L\'od\'z, PL-90236 Lodz, Poland
\and Deutsches Elektronen-Synchrotron (DESY), D-15738 Zeuthen, Germany
\and Institut de Fisica d'Altes Energies (IFAE), The Barcelona Institute of Science and Technology, Campus UAB, 08193 Bellaterra (Barcelona), Spain
\and Universit\`a  di Siena, and INFN Pisa, I-53100 Siena, Italy
\and Institute for Space Sciences (CSIC/IEEC), E-08193 Barcelona, Spain
\and Technische Universit\"at Dortmund, D-44221 Dortmund, Germany
\and Universit\"at W\"urzburg, D-97074 W\"urzburg, Germany
\and Finnish MAGIC Consortium, Tuorla Observatory, University of Turku and Astronomy Division, University of Oulu, Finland
\and Unitat de F\'isica de les Radiacions, Departament de F\'isica, and CERES-IEEC, Universitat Aut\`onoma de Barcelona, E-08193 Bellaterra, Spain
\and Universitat de Barcelona, ICC, IEEC-UB, E-08028 Barcelona, Spain
\and Japanese MAGIC Consortium, ICRR, The University of Tokyo, Department of Physics and Hakubi Center, Kyoto University, Tokai University, The University of Tokushima, Japan
\and Inst. for Nucl. Research and Nucl. Energy, BG-1784 Sofia, Bulgaria
\and Universit\`a di Pisa, and INFN Pisa, I-56126 Pisa, Italy
\and ICREA and Institute for Space Sciences (CSIC/IEEC), E-08193 Barcelona, Spain
\and also at the Department of Physics of Kyoto University, Japan
\and now at Centro Brasileiro de Pesquisas F\'isicas (CBPF/MCTI), R. Dr. Xavier Sigaud, 150 - Urca, Rio de Janeiro - RJ, 22290-180, Brazil
\and now at NASA Goddard Space Flight Center, Greenbelt, MD 20771, USA
\and Department of Physics and Department of Astronomy, University of Maryland, College Park, MD 20742, USA
% ...
}


\date{Received / Accepted }

\abstract{ 
{The large jet kinetic \modifPol{power} and non-thermal processes occurring in the microquasar SS~433} 
\modifPol{make this source a good candidate for a very high-energy (VHE) gamma-ray emitter}.
Gamma-ray fluxes above the sensitivity  \modifPol{limits} of current Cherenkov telescopes have been predicted for both the central \modifPol{X-ray} binary system and the interaction regions of SS~433 jets with the surrounding W50 nebula. Non-thermal emission at lower energies has been previously reported, indicating that efficient particle acceleration is taking place in the system.}
%
%
{We  explore the capability of SS~433 to emit VHE gamma rays during periods in which the expected flux attenuation due to periodic eclipses ($P_{\rm orb} \sim 13.1$~days) and precession of the circumstellar disk ($P_{\rm pre} \sim 162$\,days) periodically covering the central binary system is expected to be at its minimum. The eastern and western SS~433/W50 interaction regions are also examined using the whole data set available. We aim to constrain some theoretical models previously developed for this system with our observations.}
%
%
{We made use of dedicated observations from the Major Atmospheric Gamma Imaging Cherenkov telescopes (MAGIC) and High Energy Spectroscopic System (H.E.S.S.)  of SS~433 taken from 2006 to 2011. These observation were \modifPol{combined for the first time and accounted for a total effective observation time of 16.5~h}, which were scheduled considering the expected phases of minimum  absorption of the putative VHE emission. Gamma-ray attenuation does not affect the jet/medium interaction regions. In this case, the analysis of {a larger} data set amounting to $\sim$40--80\,h, depending on the region, was employed.}
%
%
{No evidence of VHE gamma-ray emission either from the central binary system or from the eastern/western interaction regions was found. Upper limits were computed for the combined data set.  \modifPol{ Differential fluxes from the central system are found to be $\lesssim 10^{-12}$--$10^{-13}$~TeV$^{-1}$~cm$^{-2}$~s$^{-1}$ in an energy interval ranging from $\sim$few $\times 100$~GeV to $\sim$few TeV. Integral flux limits down to $\sim 10^{-12}$--$10^{-13}$ ph \, cm$^{-2}$~s$^{-1}$ and $\sim 10^{-13}$--$10^{-14}$ ph \, cm$^{-2}$~s$^{-1}$ are obtained at 300 and 800 GeV, respectively.} Our results are used to place constraints on the particle acceleration fraction at the inner jet regions and on the physics of the jet/medium interactions.} 
%
%
{Our findings suggest that the fraction of the jet kinetic power that is transferred to \modifPol{relativistic} protons must be relatively small in SS~433, $q_{\rm p} \leq 2.5 \times 10^{-5}$, to explain the lack of TeV and neutrino emission from the central system. At the SS~433/W50 interface, the presence of magnetic fields $\gtrsim 10$~$\mu$G {is} derived assuming a synchrotron origin for the observed X-ray emission. { This also implies} the presence of high-energy electrons with $E_{\rm e^{-}}$ up to 50~TeV, preventing an efficient production of gamma-ray fluxes in these interaction regions.} 

\keywords{Gamma rays: observations  --- binaries: general --- X-rays: binaries ---  
stars: individual (SS~433)  --- ISM: jets and outflows}

\maketitle

\defcitealias{pavan11}{Paper I} \defcitealias{tomsick:2012eu}{T12}
\defcitealias{Reynoso2008b}{R08b}



\section{The SS~433/W50 system}
\label{section:introduction}

SS~433 \modifPol{ (R.A. 19$^h$ 11$^m$ 49.57$^s$, Dec. 4$^{\circ}$ 58'' 57.9')} {is} the first {binary system} containing a stellar{-mass} compact object in which relativistic jets were discovered  (\citealp{Abell1979}; \citealp{Fabian1979}). 
Located at a distance of 5.5~$\pm$~0.2\,kpc \citep{Blundell2004, Lockman2007}, SS~433 is an eclipsing X-ray binary system containing a black hole that is most likely $\sim$~10--20~M$_{\odot}$
 (\citealp{Margon1984})  orbiting a $\sim 30$ M$_{\odot}$ A3--7 supergiant star in a circular orbit with radius \,{79--86}~$R_{\odot}$ (\citealp{Fabrika2004}). SS~433 is extremely bright with a bolometric luminosity of $L_{\rm bol}\sim10^{40}$~erg~s$^{-1}$ (\citealp{Cherepashchuk2002}) peaking at ultraviolet wavelengths. The source displays the most powerful jets known in our Galaxy, with $L_{\rm jet}\gtrsim10^{39}$~erg~s$^{-1}$ \citep{Dubner1998, Margon1984, Marshall2002}, ejected at a relativistic velocity of 0.26\,\textit{c} \citep{Margon1989}.  The jets show a precessional period of $\sim$162.4 days with a half opening angle of $\theta_{\rm pre} \approx$ 21$^{\circ}$ with respect to the normal to the orbital plane, \modifPol {with precessional phase $\Psi_{\rm pre}=0$ defined as the phase with the maximum exposure of the accretion disk to the observer \citep{Fabrika1993}}. The inclination of the jets with respect to the line of sight subtends an angle of $i\approx$ 78$^{\circ}$ \citep{Eikenberry2001}. This value is, however, time dependent owing to precession. 

Both the high luminosity of SS~433 and its enormous jet power are thought to be a consequence of the persistent regime of supercritical accretion onto the compact object {via Roche lobe overflow  at a rate of $\dot{M} \sim 10^{-4}$\,M$_{\odot}$~yr$^{-1}$}. 
In addition, SS~433 is one of the only two X-ray binary systems in which the presence of baryons in their jets has been found \citep{Kotani1994}; the other system is 4U 1630-47 \citep{DiazTrigo2013}. Clouds of plasma with baryonic content propagate along ballistic trajectories to large distances without appreciable deceleration. At growing distances from the source, the collimated jets (with opening angle of $\approx$~1.2$^{\circ}$) can be distinguished in the X-ray, optical, and radio bands. 
The X-ray jets give rise to lines of highly ionized heavy elements \citep{Kotani1994, Marshall2002}. The emission is produced by hot gas ($T\sim 10^{8}$ K), which cools due to expansion and radiative losses while propagating outwards. Plasma at $T\sim 10^{7}$\,K is, however, still observed at large distances along the jet, indicating that a continuous source of heating is required to maintain the observed emission \citep{Migliari2002}. At radio wavelengths, the observed synchrotron flux density is about 1 Jy at 1 GHz with luminosities reaching $\sim$~4~$\times~$10$^{32}$\,erg~s$^{-1}$.

SS~433 is surrounded by the radio shell of W50, which is a large 2$^{\circ} \times 1^{\circ}$ nebula catalogued as SNR~G39.7$-$2.0 \citep{Green2006}. Its present morphology is thought to be the result of the interaction between the jets of SS~433 and the surrounding medium (\citealp{Goodall2011}). This scenario is supported by the position of SS~433 at the centre of W50, the elongation of the nebula in the east-west direction along the axis of precession of the jets (forming the so-called ``ears'' of W50; see \citealp{Safi-Harb1997}), the presence of radio, IR, optical and X-ray emitting regions also aligned with the jet precession axis, and the structure of the magnetic field through the observation of linearly polarized radio emission in the SS~433/W50 region \citep{Farnes2016}. At a distance of $\sim$10$^{20}$ cm, or about $\sim 30$\,pc, the outflowing jets are decelerated and, together with an enhanced intensity of the emission at radio wavelengths, large-scale X-ray lobes are observed. The extended X-ray emission is mostly of non-thermal origin and generally much softer than the emission from the central source (hard X-ray emission is found however in the eastern interaction regions; see \citealp{Safi-Harb1997}). At the position of the maximum of the X-ray extended emission, optical filaments perpendicular to the jet precession axis are found. Spectral analysis shows that these filaments are formed by the sweeping up of the interstellar gas and display a proper motion of 50--90\,km s$^{-1}$ (\citealp{Zealey1980, Fabrika2004}). 
At larger distances, beyond the W50 length scales, the presence of molecular clouds aligned in the direction of SS~433 jets has been reported \citep{Yamamoto2008}. These clouds, which extend for $\sim 250$\,pc at a distance of 5\,kpc, may have formed through the interaction of SS~433 jets with the interstellar H\,I gas, which would imply that SS~433 jets are more extended by a factor of $\sim 3$ than the observed X-ray jets \citep{Yamamoto2008}.

\section{Gamma-ray emission and absorption processes in SS~433/W50}

SS~433 is an exceptional laboratory to test theoretical predictions of high (100 MeV < E < 100\,GeV) and very high-energy (VHE; E > 100\,GeV) emission produced in microquasar jets { (see e.g. \citealp{{Levinson1996}, Atoyan1999, {Kaufman2002}, Bosch-Ramon2006, Orellana2007, Reynoso2008b, Bosch-Ramon2009})}. In a leptonic framework, gamma rays could be produced through inverse Compton (IC) scattering of ambient photon fields, which are dominated in this case by the supergiant companion star and {the UV and mid-IR} emission from the extended accretion disk (\citealp{Gies2002}; \citealp{Fuchs2006}). In addition, synchrotron-self Compton emission and the interaction of accelerated electrons with jet ions through relativistic Bremsstrahlung processes could also generate VHE fluxes {(\citealp{Aharonian1998})}. In a hadronic scenario, interactions of relativistic protons in the jet produce gamma rays through $\pi^{0}$ decay (see e.g. \citealp{Reynoso2008b} for a detailed study applied to SS~433). The target ions could be provided both by the companion and disk winds or by the pool of thermal protons outflowing within the jets.

Gamma rays, when produced in the inner regions of SS~433, can be strongly attenuated (see e.g. \citealp{Reynoso2008a}). Both the donor star and compact object are thought to be embedded in a thick extended envelope \citep{Zwitter1991}, \modifPol{which forms as a result of the supercritical accretion rate onto the compact object}, and provides a dense low-energy UV and mid-IR radiation field in which VHE photons are absorbed. In addition, about 10$^{-4}$\,$M_{\odot}$ yr$^{-1}$ are expelled within the $\sim$~30$^{\circ}$ half opening angle subtended by the envelope { \citep{Fabrika2008}}. {Therefore}, absorption of VHE gamma rays can also occur due to gamma-nucleon interactions through photo-pion and photo-pair production processes, whereas the photon field of the companion can also effectively reduce the gamma-ray flux through pair creation \citep{Reynoso2008a}. Such strong absorption is expected to occur along $\sim$ 80\% of the SS~433 precession cycle with a maximum at {precession phase} $\Psi_{\rm pre} \approx$~0.5, and during the regular eclipses of the compact object by the donor star at {orbital phases $\phi_{\rm orb} \approx$~0}.

The interaction regions between the SS~433 jets and the surrounding W50 nebula could also produce VHE gamma-ray emission, for example through IC scattering of cosmic microwave background (CMB) photons by electrons accelerated at the eastern ({\it{e}}1, {\it{e}}2, {\it{e}}3) and western ({\it{w}}1, {\it{w}}2)  termination regions (\citealp{Safi-Harb1997, Aharonian1998, Bordas2009}), or through {\it{pp}} interactions if protons are efficiently accelerated at the interaction shocks (\citealp{Heinz2002, Bosch-Ramon2005}). Non-thermal emission from these regions has indeed been observed from radio to X-ray energies (Brinkmann et al. 1996, \citealp{Dubner1998, Safi-Harb1997, Safi-Harb1999, Fuchs2002, Brinkmann2007}. 

Recently, high-energy gamma-ray emission from a source associated with SS~433/W50 has been reported from the analysis of $\sim$ 5 years of {\it{Fermi}}-LAT archival data  \citep{Bordas2015}. The relatively large point spread function (PSF) of the {\it{Fermi}}-LAT at the sub-GeV energies in which the source is detected, larger than $\sim 1.5^{\circ}$, prevents from an accurate localization of the emitter. 
At VHEs, SS~433/W50 remains so far undetected. Upper limits have been reported by the CANGAROO-II and HEGRA Collaborations, at the level of 2--3\% of the Crab nebula flux above 800\,GeV (\citealp{Hayashi2009, HEGRA05}), following extended observation campaigns including both the central system and the jet/medium interaction regions. The selected dates of these observations, however, did not account for the gamma-ray absorption affecting the inner system. Such constraints were instead accounted for in observations of SS~433 performed by VERITAS in 2007 and MAGIC in 2008, for which upper limits are reported in \cite{Saito2009} and \cite{Guenette2009}, respectively.

\vspace{0.5cm}
\noindent In this work, a search for VHE emission from the microquasar SS~433 with the Major Atmospheric Gamma Imaging Cherenkov telescopes (MAGIC) and High Energy Spectroscopic System (H.E.S.S.) {Imaging Atmospheric Cherenkov Telescopes (IACTs)} is reported, following dedicated observations of the source spanning several years and taken at orbital/precession phases where gamma-ray absorption should be minimal. The SS~433/W50 interaction regions are also investigated, for which a wider data set is used that is not restricted to the low-absorption phases criterion applied to the study of the inner system. The observations and analysis results are described in Section \ref{Sect:Observations} and are later discussed in Section~\ref{Sect:Discussion}.

\section{VHE observations, analysis, and results} 
\label{Sect:Observations}

\subsection{H.E.S.S. and MAGIC observations} 

\begin{table*}[t!]\footnotesize
\begin{center}
 \caption{Observations of \ssbin\ performed by H.E.S.S. and MAGIC telescopes, including the date of the observations, telescope configuration or number of operating telescopes, zenith angle range, observation live-time, and corresponding precessional phase (based on ephemeris by \citealp{Goranskij2011}).  
\label{tab:observations}}
\begin{tabular}{c c c c c c}
\hline
\hline
\modifPol{Instrument} & Epoch &  Zenith angle & Time & $\Psi_{\rm pre}$\\
 & & [$^{\circ}$] & [h] & \\
\hline
              H.E.S.S. & 30 May -- 5 June 2006 & 28--44 & 3.0 & 0.95--0.99 \\
               & 30 September -- 12 October 2007 &  38--46 &  3.1 & 0.96--0.04 \\
               & 3 -- 17 July 2009 &  33--54  & 0.9 & 0.92--0.01  \\
               & 9 -- 10 May 2011 & 28--37 & 2.1 & 0.07--0.08 \\

\hline
MAGIC & 20 -- 23 May 2010 &  24--29 & 5.6 & 0.90--0.92\\
             & 08 -- 10 June 2010 & 24--30 & 4.4 & 0.01--0.03\\
\hline
\end{tabular}
\end{center}
\end{table*}

Observations of SS~433/W50 were conducted with the H.E.S.S. and MAGIC Cherenkov telescope arrays. The H.E.S.S. is an IACT array located in the Khomas highland of Namibia \modifPol{(23$^{\circ}$S, 16$^{\circ}$E, 1800\,m above the sea level).} In its first phase, the system consisted of four identical 13\,m diameter imaging Cherenkov telescopes, covering a field of view (FoV) of about 5$^{\circ}$ diameter \citep{Bernloehr2003}. A major upgrade took place in 2012 with the addition of CT-5, which is  a 28\,m diameter telescope at the centre of the array. The data presented in this paper makes use of observations taken during the H.E.S.S.-I phase only {(CT1-4 configuration)}. In this configuration, H.E.S.S. is able to detect at a 5$\sigma$ statistical significance level a source with $\sim 0.7\%$ of the Crab nebula flux in 50h of observations \citep{Aharonian2006b}. The MAGIC is a stereoscopic system of two 17\,m diameter \modifAlicia{Cherenkov} telescopes located at the observatory El Roque de Los Muchachos (28$^\circ$N, 18$^\circ$W, 2200\,m above the sea level) \modifPol{on} La Palma, Canary Islands, Spain. \modifAlicia{Each telescope is composed of} a pixelized camera with a FoV of 3.5$^\circ$. {The sensitivity of \modifAlicia{MAGIC} at the time of the observations was $\sim$ 0.76$\% \pm 0.03\%$ of the Crab nebula flux in 50\,h above 290\,GeV \citep{MAGIC_performance_2012}}

The H.E.S.S. and MAGIC data set includes observations of SS~433/W50 taken in 2006, 2007, 2009, 2010, and 2011. Whereas the 2006 and 2007 H.E.S.S. observations were part of the H.E.S.S. Galactic Plane Survey (HGPS;\ \citealp{HESS2016}), both the MAGIC and H.E.S.S. campaigns in 2009, 2010, and 2011 were dedicated to \ssbin. Making use of the ephemeris provided by \cite{Goranskij2011}, the latter were scheduled at times in which the source was found at orbital and precession phases $\Psi$ where absorption of its putative VHE emission is expected to be at its minimum, {$\Psi_{\rm pre}$ =  0.9 -- 0.1} \citep{Reynoso2008a}. The total H.E.S.S. effective exposure time on SS~433/W50 amounts to 45h of data after standard quality selection cuts. A total effective exposure time of 8.7~h is available for the central system after selecting low-absorption precession/orbital phases. In May and June 2010, MAGIC performed observations of \ssbin\ {in stereo mode} for 10~h. After quality cuts, 7.8\,h of good data remained. A summary of the H.E.S.S. and MAGIC observations of \ssbin\ is collected in Table~\ref{tab:observations}.

The H.E.S.S. observations of the central system were performed at zenith angles ranging between 28$^{\circ}$ to 54$^{\circ}$, with an average of 34$^{\circ}$, while MAGIC observed at zenith angles between 24$^\circ$ and 30$^\circ$. The observations in both H.E.S.S and MAGIC were performed {in} wobble mode \citep{Fomin1994}, {with an offset of} $0.4^{\circ}$ for MAGIC and $0.7^{\circ}$ for H.E.S.S., respectively,  away from the source position to simultaneously evaluate the background. This observation mode allowed imaging of not only the central binary system, but also the eastern ({\it{e}}1, {\it{e}}2, {\it{e}}3) and western ({\it{w}}1, {\it{w}}2) interaction regions with the W50 nebula. In the case of H.E.S.S., the interaction regions  were also observed as part of the HGPS programme { \citep{Aharonian2006a, HESS2016}},  at zenith angles between 25$^{\circ}$ to 57$^{\circ}$, with an average of  35$^{\circ}$--38$^{\circ}$ depending on the region. The total exposure time varies from region to region (see Table~\ref{tab:SS433_INT_ULs}).

\subsection{Analysis} 

Data analysis was performed following the standard analysis procedure for each of the two {instruments} (see \citealp{Aharonian2006b} for H.E.S.S. and {\citealp{performance2016}} for MAGIC analysis details). The imaging technique is based on the parameterization of the images formed in the camera plane in order to extract the information contained in the shower with the Hillas parameters \citep{Hillas1985}. The signal extraction was performed by the reconstruction and calibration of the size and arrival time of the Cherenkov pulses. The event reconstruction was obtained by image cleaning and shower parameterization, whereas the signal and background discrimination and energy estimation were obtained by comparison of the Hillas parameters with look-up tables for a given shower intensity and impact distance (see \citealp{Aharonian2006b,MAGIC_performance_2012}), or by training an algorithm to perform gamma/hadron separation via the random forest (RF) method \citep{Albert2008b}. The event direction was derived in stereoscopic observations from the intersection of the major axes of the shower images in multiple cameras. Finally, the signal was extracted geometrically from the angular distance {$\theta^2$} ; i.e. the angular distance between the source position and the estimated source position for a given event. The signal is then determined by an upper cut in these angles, since gamma rays are reconstructed with small angles and the background follows a featureless, almost-flat distribution. For the H.E.S.S. analysis, an independent cross-check with the {\it{model analysis}} technique \citep{Naurois2009} was performed, making use of an independent calibration procedure of the raw data, with both the analysis chains providing compatible results. {\it{Standard cuts}} were used, where a cut of 60 photoelectrons on the intensity of the extensive air showers is applied, providing a mean energy threshold of $\sim 287$\,GeV for the analysis reported here. The energy threshold reached by MAGIC is 150\,GeV. A point-like source was assumed for the analysis of \ssbin. The interaction regions display extended emission at lower energies. To account for such extension, the MAGIC and H.E.S.S. analyses were optimized assuming a source radius ($\theta$-cut) of 0.05$\deg$, 0.17$\deg$, 0.25$\deg$, 0.07$\deg$, and 0.07$\deg$ for  {\it{e}}1, {\it{e}}2, {\it{e}}3, {\it{w}}1, and {\it{w}}2, respectively, derived from the extension of these regions observed at X-ray energies (see e.g. \citealp{Safi-Harb1997,  Safi-Harb1999, HEGRA05}, and references therein). 

\subsection{Results} 
\label{results}

\begin{figure}[t]
\centering
\includegraphics{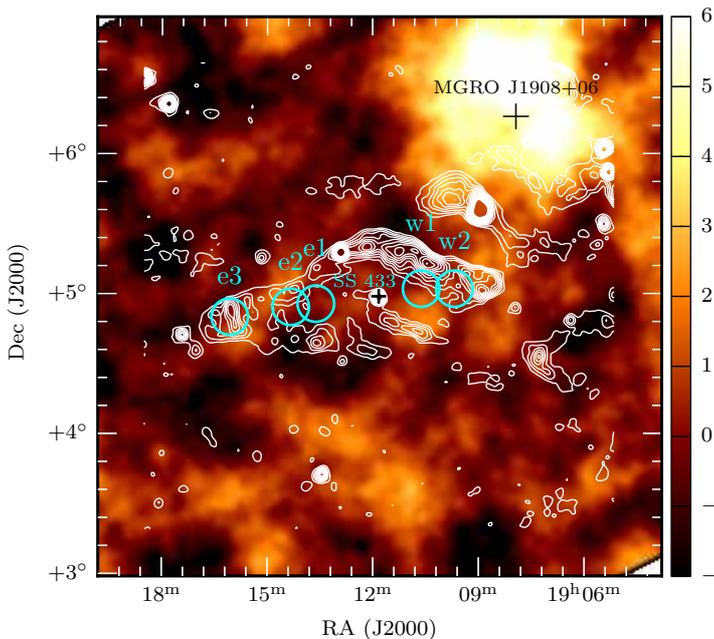}
\caption{Significance map, derived from the H.E.S.S. data, for the FoV { centred at the position of SS~433/W50} at $E\ge$~287\,GeV. 
GB6 4.85~GHz radio contours (white, from \citealt{GB1996}) 
are \modifPol{superimposed}. Cyan circles indicate the positions of the interaction regions \textit{e1, e2, e3} (eastern ``ear") and \textit{w1, w2} (western ``ear").
The bright source located north-west of SS~433 is MGRO J1908+06 \citep{Abdo2007}.
}
\label{fig:SS433_wobble_skymap2}
\end{figure}

The H.E.S.S. and MAGIC observations reported here do not show any significant VHE emission either for the central source \ssbin\ or for any of the interaction regions with the W50 nebula {\it{e}}1, {\it{e}}2, {\it{e}}3, {\it{w}}1, and {\it{w}}2 (see Fig.~\ref{fig:SS433_wobble_skymap2}). Integral upper limits \citet{Rolke2005} have been calculated at $E \geq$ 300\,GeV and at $E \geq$ 800\,GeV; the latter allows for a comparison with previous results on the source reported by the HEGRA \citep{HEGRA05} Collaboration. The results are summarised in Table~\ref{tab:SS433_INT_ULs}.  A day-by-day analysis of the H.E.S.S. and MAGIC data sets was also performed without any signature of significant emission, which could suggest a flaring episode during the dates of observation.

\begin{figure}[t!]
\centering
\includegraphics[width=\columnwidth]{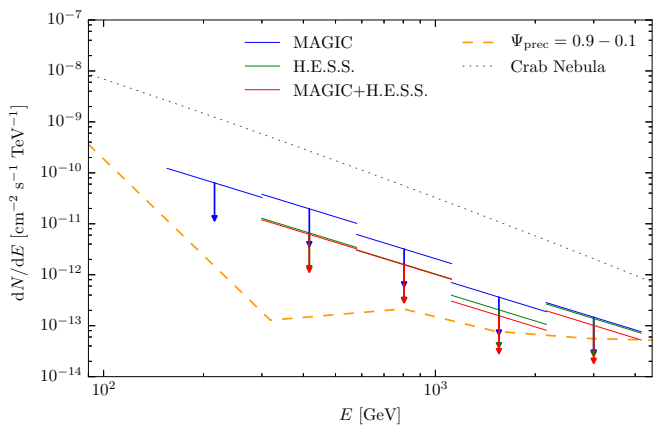}
\caption{Differential flux upper limits (95\% C.L.) from \ssbin\ obtained with MAGIC (blue), H.E.S.S. (green) and a combination of both telescopes (red) assuming a power-law with a spectral index $\Gamma = 2.7$ for the differential gamma-ray flux. The predicted differential gamma-ray flux from \cite{Reynoso2008b} for precessional phases $\Psi_{\rm pre} \in [0.9,0.1]$ in which absorption of VHE emission should be at its lower level is also displayed (dashed orange), together with the Crab Nebula flux, for reference (from \citealt{2012A&A...540A..69A}).}
\label{diff_ul_SS4332}
\end{figure}

The H.E.S.S. and MAGIC observations were used to compute the differential flux upper limits for the VHE emission from the central binary system at orbital/precession phases where absorption should be at its minimum. These limits were computed through a maximum-likelihood ratio test applied to the combined data sets obtained by both observatories. \modifPol {Events in the signal region ($n_{\rm ON}$) and in the background control regions ($n_{\rm OFF}$) from each instrument are collected in addition to the ratio of the areas in the signal and the background regions  ($\alpha$), effective area ($A_{\rm eff}$) and effective observing time $t_{\rm eff}$ corresponding to the observations of each instrument. A likelihood profile is then computed in each studied energy bin ($\Delta E_{\rm i}$) for both the signal and background distributions. Systematic uncertainties are accounted for through the inclusion of additional likelihood profiles for the distributions of $\alpha$, $A_{\rm eff}$, and energy resolution, assuming systematics at the level of $\delta \alpha = 10\%$, $\delta A_{\rm eff} = 15\%$, and $\delta E_{\rm i} = 15\%$ for the measurements of these quantities by each instrument \citep{Aharonian2006b, performance2016}. The inclusion of these systematics results in an enhancement by $\sim $15\% to 30\% on the final combined differential flux upper limit values, depending on the studied energy bin. To obtain the final combined differential flux upper limits, a likelihood ratio test is employed assuming a given range of values for the normalization factor of the gamma-ray differential spectrum, $N_{0}$. From the maximum of the likelihood profile, a 95\% confidence interval for the differential upper limit in each energy bin $\Delta E_{\rm i}$ is derived through $dN/dE = N_{0} \times E^{-\Gamma}$, where a fixed spectral index $\Gamma = 2.7$ was assumed. The final differential upper limits are shown in Fig.~\ref{diff_ul_SS4332}, both for each instrument and the combined values, together with the Crab nebula flux, for reference, and the theoretical predictions on the gamma-ray flux from SS433 expected at low-absorption precession phases $\Psi \in [0.9, 0.1]$ by \cite{Reynoso2008b}.
}

\begin{table}[ht!] \footnotesize
\begin{center}
\caption{Integral H.E.S.S. and MAGIC flux upper limits derived for \ssbin\ during low-absorption orbital/precessional phases and for the eastern/western interaction regions indicated in Fig. 1 using all available data. The results obtained with HEGRA (\citealp{HEGRA05}) are also included for comparison. Columns denote from left to right: the region of study (with coordinates and extension radius for the interaction regions) IACT instrument, effective exposure time, energy threshold for the UL calculation, and integral flux UL computed at 99\% C.L.\,\label{tab:SS433_INT_ULs}}
\scalebox{0.95}[0.95]{
 \begin{tabular}{ccccc}
 \hline
\hline
\multirow{2}{*}{Region}  &  \multirow{2}{*}{IACT} & $t_{\rm eff}$ & 300 GeV UL & 800 GeV UL \\
&  & [h] & [cm$^{-2}$~s$^{-1}$] & [cm$^{-2}$~s$^{-1}$] \\
\hline
\hline
SS~433 &  HEGRA & 96.3 & -- & 8.9 $\times 10^{-13}$  \\
 \multirow{2}{*}{RA = 19h 11m 50s} & \multirow{2}{*}{H.E.S.S.} & \multirow{2}{*}{8.7} & \multirow{2}{*}{2.3 $\times 10^{-12}$} & \multirow{2}{*}{3.9 $\times 10^{-13}$} \\
 &  &  &  &  \\
Dec =  04$^\circ$ 58' 58'' &  MAGIC & 7.8 & 1.8 $\times 10^{-12}$ & 4.3 $\times 10^{-13}$ \\

\hline
e1 & HEGRA & 72.0 & -- & 6.2 $\times 10^{-13}$  \\
RA = 19h 13m 37s     &  \multirow{2}{*}{H.E.S.S.} & \multirow{2}{*}{36.5} & \multirow{2}{*}{6.8 $\times 10^{-13}$} & \multirow{2}{*}{1.4 $\times 10^{-13}$} \\
Dec = 04$^\circ$ 55' 48''      &   &  &  &   \\
(r =  0.05$^\circ$)    &  MAGIC & 7.8 & 1.6 $\times 10^{-11}$ & 1.9 $\times 10^{-12}$  \\

\hline
e2  & HEGRA  & 73.1 & -- & 9.2 $\times 10^{-13}$  \\
RA = 19h 14m 20s &  \multirow{2}{*}{H.E.S.S.} & \multirow{2}{*}{34.8} & \multirow{2}{*}{6.0 $\times 10^{-13}$} & \multirow{2}{*}{1.3 $\times 10^{-13}$} \\
Dec =  04$^\circ$ 54' 25''  &  &  &  &  \\
(r=  0.17$^\circ$)   &  MAGIC & 7.8 & 1.7 $\times 10^{-11}$ & 2.0 $\times 10^{-12}$ \\

\hline
e3 &  HEGRA  & 68.8 & -- & 9.0 $\times 10^{-13}$  \\
RA = 19h 16m 04s   &  \multirow{2}{*}{H.E.S.S.} & \multirow{2}{*}{18.9} & \multirow{2}{*}{1.1 $\times 10^{-12}$} & \multirow{2}{*}{9.3 $\times 10^{-13}$} \\
Dec =  04$^\circ$ 50' 13''  &  & &  &   \\
(r = 0.25$^\circ$)   &  MAGIC & 7.8 & 8.7 $\times10^{-12}$ & 6.1 $\times 10^{-13}$\\

\hline
w1 & HEGRA  & 104.9 & -- & 6.7 $\times 10^{-13}$  \\
RA = 19h 10m 37s  &  \multirow{2}{*}{H.E.S.S.}& \multirow{2}{*}{62.5} & \multirow{2}{*}{2.2 $\times 10^{-13}$} & \multirow{2}{*}{4.0 $\times 10^{-14}$} \\
Dec =  05$^\circ$ 02' 13''  &  &  &  &   \\
(r = 0.07$^\circ$)   &  MAGIC & 7.8 & 1.3 $\times 10^{-11}$ & 2.2 $\times 10^{-12}$ \\

\hline
w2 &  HEGRA  & 100.7 & -- & 9.0 $\times 10^{-13}$  \\
RA = 19h 09m 40s  &  \multirow{2}{*}{H.E.S.S.}& \multirow{2}{*}{60.8} & \multirow{2}{*}{3.2 $\times 10^{-13}$ } & \multirow{2}{*}{7.6 $\times 10^{-14}$}\\
Dec =  05$^\circ$ 02' 13''  &  &  &  &   \\
(r = 0.07$^\circ$)  &  MAGIC & 7.8 & 1.4 $\times 10^{-11}$ & 2.6 $\times 10^{-12}$ \\

\hline
 \end{tabular}}
 \end{center}
 \end{table}

\section{Discussion}
\label{Sect:Discussion}

The H.E.S.S. and MAGIC observations reported here do not show any significant signal of VHE emission from SS~433/W50. The variable absorption of a putative VHE gamma-ray flux emitted from the inner regions of the binary system, which could be responsible for this non-detection, is accounted for in this study by selecting observations corresponding to precession/orbital phases where this absorption should be at its minimum. The combination of the MAGIC and H.E.S.S. observations in addition provides a \modifPol{relatively wide} coverage of the relevant precession phases from 2006 to 2011. If a long-term super-orbital variability exists in SS~433 with timescales of $\sim$ few years, for example related to a varying jet injection power or the changing conditions of the absorber in the surroundings of the central compact object, such variability does not result in an enhancement of the TeV flux up to the detection level of current IACTs. 

While SS~433 {remains} undetected at VHE, the system displays non-thermal emission at lower energies along the jets and/or at the SS~433/W50 interaction regions, which ensures the presence of an emitting population of {relativistic} particles in the system. In particular for the eastern nebula interaction sites, the observed synchrotron X-ray emission implies the presence of  for example up to multi-TeV electrons in these regions \citep{Safi-Harb1999}. 

By considering in detail the photon and matter fields both from the companion star and accretion/circumstellar disks, gamma-ray fluxes from SS~433/W50 have been predicted at a level of  $\sim 10^{-12}$--$10^{-13}$\,ph cm$^{-2}$ s$^{-1}$ (see e.g. \citealp{Band_Grindlay-1989, Aharonian1998, Reynoso2008b}). \cite{Reynoso2008b} consider in particular \textit{pp} interactions between relativistic and cold protons in SS~433 jets during low-absorption precession/orbital phases, producing gamma-ray fluxes at $E_{\gamma} \ge$ 800\,GeV during these precession phases at a level of $\Phi_{\rm VHE} \approx 2.1 \times 10^{-12}$\,ph~cm$^{-2}$~s$^{-1}$.
\modifPol{The general framework used to derive the relativistic proton distribution in \cite{Reynoso2008b} has been revised by \cite{Torres2011}, who report significant deviations of these proton fluxes for jets displaying large Lorentz factors and/or small viewing angles, for example {blazar} jets and gamma-ray bursts. In SS~433, with a  moderate jet Lorentz factor of 1.036 ($v$ = 0.26\,\textit{c}; \citealp{Abell1979}) and a relatively large jet viewing angle, $\sim 78^{\circ}$ \citep{Eikenberry2001}, the correction factor on the fluxes predicted by \cite{Reynoso2008b} could be affected at the level of $\sim$ 20\%}.
The gamma-ray flux predicted by Reynoso et al. (2008) depends on the efficiency in transferring jet kinetic energy to the relativistic proton population, $q_{p} $, which is treated in their model \modifPol{as a free parameter}.
Using the HEGRA upper limits to the VHE gamma-ray flux from SS~433, $q_{p}$ is constrained  to be $\leq 2.9\, \times \, 10^{-4}$. This upper limit is obtained however under the assumption that the HEGRA observations took place during a complete precessional cycle. With the H.E.S.S. and MAGIC upper limits reported here, a more stringent constraint on the fraction of power carried by relativistic protons in the SS~433 jets is obtained, $q_{p}\lesssim \, 2.5 \, \times 10^{-5}$. 
%

In a hadronic scenario, gamma rays from $\pi^{0}$-decay should also be accompanied by \modifPol{neutrinos from the decay of charged pions}. The IceCube Collaboration { analyzed the data around the position of} SS~433/W50 with no significant detection of the source. An upper limit at 90\% confidence level (C.L.) on the muon neutrino flux at 1~TeV is set at $\phi_{\nu_{\mu}+\bar{\nu}_{\mu}}$= 0.65 $\times 10^{-12}$~TeV~cm$^{-2}$~s$^{-1}$  \citep{icecube2014}, using four years of data and assuming an $E^{-2}$ flux distribution. The model of \cite{Reynoso2008b} also predicts the neutrino flux emitted by this system with a neutrino differential flux at 1~TeV of $\phi_{\nu}$ = $2 \times 10^{-12}$\,cm$^{-2}$~$s^{-1}$ averaged over all precessional phases. The IceCube upper limits can be used to put a limit on $q_{\rm p}$ of $\sim 3.3 \times 10^{-5}$, which is marginally less restrictive than the value obtained with the gamma-ray observations. 
\modifPol{However, the gamma-ray and neutrino flux estimates of \cite{Reynoso2008b} are based on a proton acceleration efficiency of \modifPol{$\eta = t_{\rm acc}/ t_{\rm gyr} \sim 0.07;$  $t_{\rm acc}$ is the acceleration time and $t_{\rm gyr} = ceB/E_{\rm p}$, where $B$ is the magnetic field in the accelerator region and $E_{\rm p}$ is the proton energy. Accounting for adiabatic and radiation losses, these authors derive a maximum energy for relativistic protons of $E_{\rm p }\leq 10^{3}$~TeV}. Different values for the magnetic field, target proton densities, and/or adiabatic expansion velocities in the acceleration region would also imply variations in predicted gamma-ray and neutrino fluxes.}

At the {interaction} regions of  the jets of \ssbin\ with the surrounding W50 nebula, the X-ray spectra from the extended lobes are well represented by a power-law model (\citealp{Moldowan2005}); a synchrotron origin for this emission has been suggested (\citealp{Safi-Harb1997}; see also \citealp{Safi-Harb1999}) that would imply the presence of electrons with energies up to $\sim 50$\,TeV in those regions. The VHE gamma-ray emission from the SS~433/W50 interaction regions was first considered by \cite{Aharonian1998}, who estimated gamma-ray fluxes at a level of $\sim 10^{-12}$\,ph cm$^{-2}$ s$^{-1}$ for the eastern \textit{e3} region produced by electrons scattering off CMB photons. \cite{Bordas2009} also considered the non-thermal emission produced in microquasar jets/ISM interaction regions, providing gamma-ray flux estimates as a function of the kinetic power of the jets, age of the system, and  particle density of the environment. The application of this model to SS~433/W50 yielded fluxes at the level of $\sim $10$^{-13}$\,erg cm$^{-2}$ s$^{-1}$ for $E$ > 250\,GeV for an assumed distance to the system of 5.5\,kpc \citep{Bordas2010}, which are roughly at the level of the upper limits reported here. However, as noted in \cite{Aharonian1998} (see also discussion in \citealp{Safi-Harb1999} and \citealp{HEGRA05}), electrons accelerated at the interaction region shock interface could lose most of their energy mainly through synchrotron emission for ambient magnetic fields at or above $\sim 10\,\mu$G, preventing an effective channelling through IC scattering that is relevant for the production of gamma rays at high and very high energies. The integral flux upper limits for the interaction regions shown in Table~\ref{tab:SS433_INT_ULs}, together with the assumption that the same high-energy electron population is responsible for the observed (synchrotron) X-ray emission and the putative gamma-ray fluxes, can be used to constrain the magnetic fields present in the shocked SS~433 jets/ISM interaction regions. \modifPol{ \cite{Rowell2001} \citep[see also][]{HEGRA05} make use of HEGRA upper limits obtained for the $e$3 interaction region ($\Phi_{\rm HEGRA} \leq 2.1 \times 10^{-12}$\,ph~cm$^{-2}$~s$^{-1}$) and the predictions by \citet {Aharonian-1997} to derive a lower limit on the post-shock magnetic field in this region of $\sim 13\,\mu$G. Using the upper limits reported here, a more constraining lower limit on the magnetic field of {20--25\,$\mu$G} is obtained}. 

The huge kinetic luminosity \modifPol{and baryonic matter transported} by the SS~433 jets and the presence of the surrounding target material provided by the disk wind and/or the W50 nebula render {\it pp} interactions at those larger scales a good TeV emission mechanism as well, as shown for even more modest energy budgets (see e.g. \citealp{Bosch-Ramon2005}). \modifPere{\cite{Bordas2015} found a gamma-ray signal from the direction of SS~433/W50. {The steadiness of the flux and the  derived spectral properties, with the gamma-ray emission extending only from ~200 MeV to ~800 MeV, prompted \cite{Bordas2015} to suggest a jet-medium interaction scenario for the observed emission. A cut-off power law was needed to fit the {\it Fermi}-LAT spectrum with cut-off energies of a few GeV. The upper limits reported here are therefore fully consistent with the {\it Fermi}-LAT extrapolation of the fitted spectra. If the MeV/GeV emission is produced by relativistic particles in SS~433 jets, as suggested in \cite{Bordas2015}, the acceleration mechanism may be only relatively efficient, thereby preventing a significant detection of the system in the VHE regime.}}

The upper limits reported here for SS~433 and those obtained on the steady emission from other well-established Galactic microquasars {(e.g. Cyg X-1, Cyg X-3, GRS 1915+105, MWC 656, and Cir X-1; see e.g. \citealp{Nicholas_Rowell-2008, HESS-2009, Saito2009, Aleksic2010, Aleksic2015a, HESS_MQ_2016})} imply that if their jets are also baryon loaded as in SS~433 (see also the case of 4U1630-47, \citealp{DiazTrigo2013}), their contribution to the Galactic cosmic-ray flux must be limited to relatively low energies in the GeV domain, as more efficient proton acceleration is constrained by the lack of VHE {gamma-ray} emission. \\

\begin{acknowledgements}


The authors would like to thank M. M. Reynoso for sharing data and information from his paper \cite{Reynoso2008b}.

The MAGIC Collaboration would like to thank the Instituto de Astrof\'{\i}sica de Canarias for the excellent working conditions at the Observatorio del Roque de los Muchachos in La Palma. The financial support of the German BMBF and MPG, the Italian INFN and INAF, the Swiss National Fund SNF, the ERDF under the Spanish MINECO (FPA2015-69818-P, FPA2012-36668, FPA2015-68378-P, FPA2015-69210-C6-2-R, FPA2015-69210-C6-4-R, FPA2015-69210-C6-6-R, AYA2015-71042-P, AYA2016-76012-C3-1-P, ESP2015-71662-C2-2-P, CSD2009-00064), and the Japanese JSPS and MEXT is gratefully acknowledged. This work was also supported by the Spanish Centro de Excelencia ``Severo Ochoa'' SEV-2012-0234 and SEV-2015-0548, and Unidad de Excelencia ``Mar\'{\i}a de Maeztu'' MDM-2014-0369, by the Croatian Science Foundation (HrZZ) Project 09/176 and the University of Rijeka Project 13.12.1.3.02, by the DFG Collaborative Research Centers SFB823/C4 and SFB876/C3, and by the Polish MNiSzW grant 745/N-HESS-MAGIC/2010/0.

The support of the Namibian authorities and the University of Namibia in facilitating the construction and operation of  H.E.S.S. is gratefully acknowledged, as is the support by the German Ministry for Education and Research (BMBF), the  Max Planck Society, the German Research Foundation (DFG),  the French Ministry for Research, the CNRS-IN2P3 and the  Astroparticle Interdisciplinary Programme of the CNRS, the  U.K. Science and Technology Facilities Council (STFC), the  IPNP of the Charles University, the Czech Science Foundation,  the Polish Ministry of Science and Higher Education, the South  African Department of Science and Technology and National  Research Foundation, and by the University of Namibia. We appreciate the excellent work of the technical support staff in Berlin, Durham, Hamburg, Heidelberg, Palaiseau, Paris, Saclay, and in Namibia in the construction and operation of the equipment.

\end{acknowledgements}


\bibliographystyle{aa} \bibliography{SS433_aa}

\end{document}